\documentclass[a4paper]{jpconf}
\usepackage{graphicx}
\begin{document}
\title{Modern Particle Physics Event Generation with WHIZARD}

\author{J Reuter$^1$, F Bach$^1$, B Chokouf\'{e}$^1$, W Kilian$^2$, T
  Ohl$^3$, M Sekulla$^2$ and C Weiss$^{1,2}$}

\address{$^1$DESY Theory Group, Notkestr. 85, D--22607 Hamburg, Germany}
\address{$^2$University of Siegen, Department of Physics,
  Walter-Flex-Str. 3, D--57068 Siegen, Germany}
\address{$^3$University of W\"urzburg, Department of Physics and
  Astronomy, Emil-Hilb-Weg 22, D--97074 W\"urzburg}

\ead{juergen.reuter@desy.de, fabian.bach@desy.de,
  bijan.chokoufe@desy.de, kilian@hep.physik.uni-siegen.de,
  ohl@physik.uni-wuerzburg.de, sekulla@hep.physik.uni-siegen.de,
  christian.weiss@desy.de}

\begin{abstract}
We describe the multi-purpose Monte-Carlo event generator WHIZARD 
for the simulation of high-energy particle physics experiments. Besides
the presentation of the general features of the program like SM physics,
BSM physics, and QCD effects, special emphasis will be given to the support
of the most accurate simulation of the collider environments at hadron 
colliders and especially at future linear lepton colliders. On the more 
technical side, the very recent code refactoring towards a completely 
object-oriented software package to improve maintainability, flexibility 
and code development will be discussed. Finally, we present ongoing work
and future plans regarding higher-order corrections, more general model
support including the setup to search for new physics in vector boson
scattering at the LHC, as well as several lines of performance
improvements. 
\end{abstract}

\section{Introduction}

\texttt{WHIZARD}~\cite{Kilian:2007gr} is a general event generator
for all kinds of scattering and decay processes at high-energy
hadron and lepton colliders.  

The default matrix element generator of \texttt{WHIZARD} is
\texttt{O'Mega}~\cite{Moretti:2001zz}. This latter subpackage provides
matrix elements for multi-leg tree-level processes, using the 
helicity formalism. The high-dimensional phase-space integrations are
performed by the multi-channel Monte-Carlo integrator
\texttt{VAMP}~\cite{Ohl:1998jn}. Its algorithm is
adaptive both between and within channels, and thus computes accurate
phase-space integrals and efficiently generates weighted and
unweighted event samples. 

The \texttt{WHIZARD} core acts as a connector of these different
components. The core contains the algorithm for multi-channel 
phase-space parameterization and mapping, provides the user
interface, and also interfaces external 
programs (e.g., parton distributions, event formats, hadronization),
the routines for writing and reading event files, and modules for
parton shower and jet physics. Further modules allow for the 
for numerical analyses and visualization of event samples.

In order to be able to describe realistic ILC and CLIC environments, 
\texttt{WHIZARD} contains a dedicated package for beam-spectrum
simulation, \texttt{CIRCE}~\cite{Ohl:1996fi}. As an alternative, 
\texttt{GuineaPig} beam-event samples can be fed into \texttt{WHIZARD}
using \texttt{CIRCE2}.

\section{Program Overview}

\texttt{WHIZARD} -- initiated along the lines of the TESLA design
study~\cite{Kilian:2001qz} to provide improved simulations for
electroweak processes at lepton colliders -- has developed into a
universal generator for partonic events at all types of present and
future (lepton and hadron) colliders. 

With the new millenium, the implementation of QCD color flows
finalized the support for the full Standard Model (SM), and parton
distributions, different event samples and an ever enlarging list of
models beyond the SM (BSM) came up. Here, we should mention
particularly the implementation of the MSSM and several variations
thereof, like the NMSSM~\cite{Reuter:2009ex}. For this purpose, the
SUSY Les Houches Accord
(SLHA)~\cite{Skands:2003cj,Allanach:2008qq,AguilarSaavedra:2005pw} has
been provided in order to interface SUSY spectrum
generators. \texttt{WHIZARD} has been used for a plethora of both
theoretical and experimental studies for TESLA and the ILC as well as
the LHC. The SLAC event database for Linear Collider events has been
generated with the package.

To meet the enormous technical demands of event simulations in the LHC
and ILC era, the core of the program has been thoroughly rewritten in 
2007-2010, which gave rise to the \texttt{WHIZARD}~2 release series.

The rigid input files of the early versions have been replaced in
\texttt{WHIZARD}~2 by input commands within the domain-specific
language SINDARIN. Besides a simplification and unification of the
input inside the user interface, the user gets the full power of a
programming language at hand. With the help of SINDARIN, cuts,
analyses, interfaces, process collection, parameter scans and many
more things can be defined. 

By means of external packages like SARAH~\cite{Staub:2013tta} or 
\texttt{FeynRules}~\cite{Christensen:2008py}, interfaced to
\texttt{WHIZARD}~\cite{Christensen:2010wz}, new physics models can
easily be added. Inside \texttt{WHIZARD}, complete quantum-mechanical
correlations are kept using a generic density-matrix formalism. In
addition, the program is endowed with its own parton-shower
module~\cite{Kilian:2011ka} as an alternative to showering being
performed externally. Exact matrix elements can be matched to the
shower by means of the MLM scheme, while other merging and matching
schemes are planned for the near future.

\texttt{WHIZARD}~2 is written in Fortran2003 that is supported in
modern Fortran compilers. Backwards compatibility is kept until
version 4.7.4 of \texttt{gfortran}. The matrix element generator
\texttt{O'Mega} is written in the functional programming language
\texttt{OCaml}, for which complete packages exist for all major
platforms. The current \texttt{WHIZARD} production version is 2.2.2
(as of October 2014), while the next release, 2.2.3, is
planned for end of November. The complete package is thoroughly tested
and can be installed on all recent Linux and MAC OS systems.


\section{Technical Details}
\label{sec:techdetails}

The standard installation of \texttt{WHIZARD} is foreseen centrally on
a machine, but it can of course also be installed locally in a user
directory. It uses the standard toolchain of \texttt{automake},
\texttt{autoconf}, and \texttt{libtool}. The program is located at the
HepForge server~\cite{manual}. Downloads are available either as
tagged versions (\texttt{.tar.gz} format), or as a development version
from the public \texttt{svn} repository.  The package conforms to the
standard installation procedure, using the usual \texttt{configure} --
\texttt{make} -- \texttt{make install} chain and the optional
\texttt{make check} and \texttt{make installcheck} steps for
additional safety checks.

Event generation and simulation projects of the users can then be set
up in arbitrary directories without any predefined structure. For this
purpose, a single command \texttt{whizard} exists, calling a single
SINDARIN command file as input.  Alternative modes of using
\texttt{WHIZARD} exist, namely an interactive mode, or linking it as a  
subroutine library that is interoperable with C, C++ or any other
C-compatible language (e.g., Python). The package generates and
processes matrix element codes on-the-fly as dynamically linked
libraries. There is also a statically linked mode for the work e.g. on
batch clusters.

According to the modern object-oriented programming paradigm, there
should be a clear separation between abstract type declarations and
specific implementations. This is reflected in the modern Fortran
standard, and hence \texttt{WHIZARD}~2 is broken down into these
building blocks. There has been a major refactoring of the code 
between versions 2.1 and 2.2 (which is still partially ongoing) along
these lines. The consequent separation of interface from 
implementation greatly improves the maintainability and enhances 
the possibilities for future module replacements, reimplementations,
and extensions enormously. This is now realized in most parts of the
core code like the process structure, matrix element calculation
methods, beam structure, integration methods, decays, the phase space
and the shower. For parallelization, \texttt{WHIZARD}~2 uses
OpenMP. This is especially used in a new high-performance virtual
machine and will be made available in the upcoming version 2.2.3. An
MPI implementation is foreseen for the nearer future. 

Quality assignment for \texttt{WHIZARD} is provided at first by using
the \texttt{svn} version control at HepForge. Secondly, a continuous
integration system runs all commits automatically through a chain of
several hundreds of unit and function tests. Bugs and feature requests
are steered by means of HepForge's tracking system.


\section{Physics}

The main method for matrix elements inside \texttt{WHIZARD} is
provided by the matrix element generator
\texttt{O'Mega}~\cite{Moretti:2001zz} which is able to generate
complete tree-level matrix elements with multiple 
external legs (successful tests on standard hardware have involved up
to 15 external particles).  \texttt{O'Mega} is based on a recursive
algorithm that reuses \emph{all} common subexpressions and replaces
the forest of all tree diagrams by the equivalent directed acyclical
graph (DAG). It is able to provide the complete color correlations of
QCD matrix elements, using the color-flow formalism with phantom
$U(1)$ particles~\cite{Kilian:2012pz} as an efficient way to generate
colorized DAGs that can be evaluated exactly or be projected onto
color-flow amplitudes.

\texttt{WHIZARD} now supports processes to consist of several
different components, e.g. for inclusive production samples using
process containers or to combine NLO subtraction terms with real
emission and virtual matrix elements. Also available are flavor sums
which are however technically different, as masses have to be equal to
commonly use the same phase space. Within the SINDARIN steering
language this feature is easily available for the user with an
appropriate syntax for inclusive processes and a detailed
specification of decay chains. The latter will be refined in upcoming
versions. Exact spin and color correlations are kept using the
internal density-matrix formalism inside \texttt{WHIZARD}.

\texttt{WHIZARD}~2 can not only process complete matrix elements, but
also generate decay chains and cascades. They consist of arbitrarily
chosen elementary processes to be integrated each separately, but
concatenated for the event generation. The default option is to take
full spin correlation among intermediate states into account. There is
also an option to restrict to classical spin correlations (only
diagonal entries in the spin-density matrix), or to even switch off
spin correlations completely in order to test the importance of spin
correlations. \texttt{WHIZARD} can also set up decays and
branching fractions for chosen physics models automatically. 

Arbitrary factorization and renormalization scale settings (affecting
QCD) can be used in \texttt{WHIZARD}~2 using the same syntax
expressions as the one for cuts or analyses. Furthermore, since  
version \texttt{WHIZARD} 2.2 there is the possibility to
reweight existing event samples, generated either internally or read
from file, when changing the setup of the original process,
e.g. parameters in the hard matrix elements, the event scale, the
chosen structure functions, or the QCD parton shower.


\section{Linear Collider Simulation}
\label{sec:lcsim}

\texttt{WHIZARD} is an event generator for all kinds of high-energy
colliders, but a particular focus has always been its vast support for
a realistic lepton collider environment. A description as accurate as
possible of the beam properties is manadatory for studies and
analyses, given the required level of precision at ILC and CLIC.  

\texttt{WHIZARD} incorporates the \texttt{CIRCE1}~\cite{Ohl:1996fi}
package that parameterizes the beams of an $e^+e^-$ collider and
provides an event generator for factorized beam spectra. Adapting
this generator (or, alternatively, the parameterized spectrum
directly), \texttt{WHIZARD} can integrate and simulate any $e^+e^-$ process
with a realistic beam description.  As an upgrade to previous
versions, \texttt{WHIZARD}~2.2 ships with beam spectra that correspond to the
current ILC design parameters. To account for cases where such a
factorized form is insufficient, \texttt{WHIZARD} can alternatively read
beam-event files as they are produced by \texttt{GuineaPig}. From
version 2.2.3, the \texttt{CIRCE2} package inside \texttt{WHIZARD} now
directly interfaces \texttt{GuineaPig(++)} and allows for the use of
correlated lepton collider beam spectra with their steeply rising
peaks where a power-law parameterization is insufficient.


Lepton-collider processes are not only influenced by beamstrahlung,
but also strongly affected by electromagnetic initial-state radiation
(ISR). \texttt{WHIZARD} uses a standard structure-function formalism
that resums the corrections from infrared (leading) and collinear (3rd
order) radiation to implement ISR. It takes ISR into account
both in kinematics and dynamics, if requested.

\texttt{WHIZARD} enables the user to specify arbitrary beam
polarizations, ranging from unpolarized, left- or right-handed
circular or transversal polarization to arbitrary spin-density
matrices. The polarization and polarization fractions are specified
for both beams independently. The user can also define asymmetric beam
setups and a crossing angle, which will be taken into account in the
kinematics setup.

Photons as initial particles are available in various incarnations:
on-shell, radiated from $e^\pm$ (effective photon approximation), or
beamstrahlung photons generated by \texttt{CIRCE1}.  A photon-collider
option that uses the \texttt{CIRCE2} beam description is also available but
no longer maintained, due to the lack of current ILC or CLIC
photon-collider mode beam parameters and simulations of the
corresponding beam-beam interactions.

For cuts, reweighting and internal analysis, \texttt{WHIZARD} employs its
dedicated language SINDARIN that allows for computing a wide range of
event-specific and generic observables.

\texttt{WHIZARD} supports various event output formats, including the
traditional HEPEVT format and its derivatives, StdHEP, LHEF, HepMC,
and others.  A direct interface to LCIO is planned.


\section{QCD}
\label{sec:qcd}

For a precise calculation of exclusive processes, a
program needs fine control over QCD corrections.  Regarding real
radiation, the multi-leg capability of \texttt{WHIZARD} allows to include
high orders of the QCD coupling.  Collinear and soft radiation in
exclusive events is affected by large
logarithms, which are conveniently resummed in the semi-classical
approach of a parton shower algorithm.

Using standard event formats and suitable cuts, \texttt{WHIZARD} allows for
attaching an external parton-shower generator.  \texttt{WHIZARD}~2
furthermore contains an internal showering module in two different
incarnations: a $k_T$ ordered shower along the lines of the \texttt{Pythia}
shower~\cite{Sjostrand:2006za}, and an analytic parton shower, which
keeps the complete shower history and allows to reweight
it~\cite{Kilian:2011ka}.  There is support for combining exact matrix
elements and QCD radiation from the parton shower using the MLM
matching prescription.  These modules are foreseen to receive a more
detailed validation, tuning and further improvements after the 2.2
release.

Beyond the parton shower, hadronization and hadronic decays
are not performed by \texttt{WHIZARD} internally, but can be applied to the
generated partonic event samples via the \texttt{Pythia}~6
package~\cite{Sjostrand:2006za} which is attached to the \texttt{WHIZARD}
distribution, or using e.g. LHE event samples that are then fed into a
external (shower and) hadronization package.

A high-luminosity linear collider will be capable of a high-precision
scan of the top-quark pair-production threshold~\cite{ILCTDR}.  To
match this on the theoretical side, one needs to resum logarithms of
the top velocity $\sim\alpha_s\ln v$ as well as gluon Coulomb
potential terms $\sim\alpha_s/v$ in a non-relativistic approach and to
relate this to the relativistic matrix elements in the continuum.
There is an ongoing project for including these effects in
\texttt{WHIZARD}~2 which will make the theoretical calculation
available in the simulation of exclusive events.  As a first step, the
next-to-leading-logarithmic approximation matched to
next-to-leading-order matrix elements has already been
implemented and will be included in the upcoming
release 2.2.3~\cite{topthreshold}.


\section{Status of next-to-leading order calculations}

\texttt{WHIZARD}~2 with \texttt{O'Mega} matrix elements is an event
generator of tree-level processes.  There have been several projects
that extended it to next-to-leading order, including loop corrections
and proper infrared-collinear subtraction.
Ref.~\cite{Robens:2008sa,Kilian:2006cj} describes the extension of
\texttt{WHIZARD}~1 to a positive-definite NLO event generator for the
electroweak pair production of charginos in the MSSM, including full
electroweak SUSY corrections matched to the photon initial and final
state radiation.  Independently,
Ref.~\cite{Binoth:2009rv,Greiner:2011mp} implemented the QCD NLO
correction with subtraction for a particular LHC process.  Along these
lines, the Binoth Les Houches Accord (BLHA)
interface~\cite{Binoth:2010xt,Alioli:2013nda} has 
been implemented for reading and writing contract files with one-loop
programs (OLP), and has been validated.

Building upon the new data structures of \texttt{WHIZARD}~2.2, an
implementation of automatic NLO QCD corrections is currently being
developed.

\texttt{WHIZARD} is being extended to calculate cross sections at
next-to-leading order in $\alpha_s$. For this purpose, we have
implemented the FKS subtraction scheme~\cite{Frederix:2009yq}, which
relies on a partition of the phase space into disjoint regions. Using
this scheme, \texttt{WHIZARD} computes the real-subtracted and
virtual-subtracted part of the cross section. 

The virtual amplitude is calculated using
\texttt{GoSam}~\cite{Cullen:2014yla}, which can be interfaced very
easily with \texttt{WHIZARD} using the BLHA
conventions~\cite{Binoth:2010xt}.  

For the real amplitude, standard \texttt{O'Mega} matrix elements are
used. The subtraction terms require color- and spin-correlated Born matrix
elements. However, the treatment of the latter has not been adressed
yet, because spin-correlated matrix elements are only non-zero if at
least one external particle is a gluon. For the color-correlated Born
matrix element we use the usual Born matrix element multiplied with
$C_F$, which is suited for 2-jet cross sections. These parts of the
calculation will be fully generalized soon, also using \texttt{GoSam}.

The implementation has been tested using the analytically known
R-ratio for 2-jet production in lepton collisions. \texttt{WHIZARD}
correctly reproduces the result $\sigma_{\rm{NLO}} =
\left(1+\alpha_s/\pi\right)\sigma_{\rm{LO}}$. It has also been tested
for the processes $e \nu_e \rightarrow q \bar{q}' Z$ and $e^+ e^-
\rightarrow q \bar{q}' l \nu_l$ and has been validated for those
processes. 

The calculation of NLO-QCD cross sections will be an experimental
feature of Release 2.2.3 and at least supports processes including 
two colored particles in the final state. 	 


\section{Physics models}
\label{sec:physics}

As a generator of hard matrix elements, \texttt{WHIZARD} has to support
various particle species and interactions.  The allowed spin
representations for particles are 0, 1, 2 (bosons) and 1/2, 3/2
(fermions), all massive or massless, both Dirac and Majorana spinors,
optionally colored (triplet or octet).  The \texttt{WHIZARD} libraries
support all Lorentz structures for interactions in the models
described below.  A completely general framework supporting all
possible Lorentz structures is under construction.

\subsubsection*{BSM Models}

Beyond the SM and its QCD and QED subsets, \texttt{WHIZARD} supports the
minimal supersymmetric Standard Model (MSSM) with different variants
and extensions.  These include models with
gravitinos~\cite{Hagiwara:2005wg,Kalinowski:2008fk,Pietsch:2012nu} and the
NMSSM~\cite{Reuter:2009ex} (see also~\cite{Reuter:2010nx}). 

Among models with strongly interacting sectors \texttt{WHIZARD} includes
Little Higgs models in different incarnations, with and without
discrete symmetries, 
cf.~\cite{Kilian:2003xt,Kilian:2004pp,Kilian:2006eh,Reuter:2012sd,Reuter:2013iya,Reuter:2014iya}.

\texttt{WHIZARD} has also been used for studying more exotic models such as
the noncommutative
SM~\cite{Ohl:2004tn,Alboteanu:2006hh,Alboteanu:2007by} (not included 
in the official 
release).  It further supports the completely general two-Higgs
doublet model (2HDM), as well as generic models containing a $Z'$
state, and extra-dimensional models like Universal Extra Dimensions
(UED). A more detailed list can be found in the \texttt{WHIZARD}
documentation~\cite{manual}.

\subsubsection*{Effective Theories}

As an alternative tool for studying deviations from the SM, \texttt{WHIZARD}
contains SM extensions with anomalous couplings, expressible as
coefficients of higher-dimensional operators in an effective theory.
Several models in \texttt{WHIZARD}'s library define either anomalous triple
and quartic gauge boson couplings, which have been used for studies at
LCs~\cite{Beyer:2006hx,Boos:1997gw} or at
LHC~\cite{Alboteanu:2008my,Kilian:2014zja}.  Anomalous
top-quark couplings are also supported~\cite{Bach:2012fb,Bach:2014zca}.

Recent interest in the physics of high-energy vector-boson scattering
has triggered the development and addition of simplified models for
strong interactions and compositeness (SSC)~\cite{Kilian:2014zja}.  In
addition to generic new degrees of freedom, they implement a
unitarization procedure that is required for extrapolating into the
energy range that will become accessible at ILC and, in particular, at
CLIC.


\section{Conclusion and Outlook}
\label{sec:outlook}

\texttt{WHIZARD} is a versatile and user-friendly tool for both SM and BSM
physics at all possible high-energy colliders. In these proceedings, we
report the recent progress in the technical development of the program
as well as new physics features. Two main points to be mentioned are
the special focus on lepton-collider beam spectra, which involves a 
detailed account of correlated beam spectra, arbitrary polarization
and interfaces to the machine simulation. Future development will be
devoted to a better description of QED radiation from both the initial
and final states at lepton colliders and their proper matching to hard
matrix elements. The second topic is the progress in the automation of
QCD higher-order processes by interfacing external one-loop programs
through the BLHA interface, the automatic generation of subtraction
terms for soft-collinear singularities in the real emission as well as
the virtual part and an automatic phase-space integration of the
different components by using the method of FKS regions.

Further plans for new features include the support for more general
Lorentz and color structures in models, a more convenient model
interface, power-counting of coupling constants in the matrix element,
further refinements in the beam description, and many technical
improvements.
     

\section*{Acknowledgments}

JRR wants to thank the organizers of ACAT 2014 for a great conference
in the wonderful and lovely city of Prague.


\bibliography{2014_ACAT_reuter}
\bibliographystyle{iopart-num}

\end{document}